\documentclass[twocolumn]{aastex701}

\usepackage{graphicx}
\usepackage{amsmath}
\usepackage{hyperref}
\usepackage{longtable}
\usepackage{booktabs}

\newcommand{\te}{t_{\rm E}}
\newcommand{\thetae}{\theta_{\rm E}}

\newcommand{\pie}{\pi_{\rm E}}

\newcommand{\dl}{D_{\rm L}}
\newcommand{\ds}{D_{\rm S}}
\def\e{{\rm E}}

\definecolor{brown}{rgb}{0.59, 0.29, 0.0}
\definecolor{darkgreen}{rgb}{0.0, 0.42, 0.24}
\definecolor{darkblue}{rgb}{0.01, 0.31, 0.59}
\definecolor{blue}{rgb}{0.0,0.0,1.0}
\definecolor{green}{rgb}{0.0,1.0,0.0}

\begin{document}

\title{OGLE-2018-BLG-0829L: A Microlensing System with a Planet Orbiting a Star Possessing a Brown-Dwarf Companion}
\shorttitle{OGLE-2018-BLG-0829L}

\author{Cheongho Han}
\altaffiliation{KMTNet Collaboration}
\affiliation{Department of Physics, Chungbuk National University, Cheongju 28644, Republic of Korea}
\email{cheongho@astroph.chungbuk.ac.kr}
\author[0000-0003-4625-8595]{Qiyue Qian}
\affiliation{Department of Astronomy, Westlake University, Hangzhou 310030, Zhejiang Province, China}
\affiliation{Department of Astronomy, Tsinghua University, Beijing 100084, China}
\email{qqy22@mails.tsinghua.edu.cn}
\author{Andrzej Udalski} 
\altaffiliation{OGLE Collaboration}
\affiliation{Astronomical Observatory, University of Warsaw, Al.~Ujazdowskie 4, 00-478 Warszawa, Poland}
\email{udalski@astrouw.edu.pl} 
\author{Chung-Uk Lee}
\altaffiliation{KMTNet Collaboration}
\affiliation{Korea Astronomy and Space Science Institute, Daejon 34055, Republic of Korea}
\email{leecu@kasi.re.kr}
\author[0000-0001-6000-3463]{Weicheng Zang}
\altaffiliation{KMTNet Collaboration}
\affiliation{Department of Astronomy, Westlake University, Hangzhou 310030, Zhejiang Province, China}
\email{zangweicheng@westlake.edu.cn}
\collaboration{14}{(Leading authors)}
\author{Michael D. Albrow}   
\affiliation{University of Canterbury, Department of Physics and Astronomy, Private Bag 4800, Christchurch 8020, New Zealand}
\email{michael.albrow@canterbury.ac.nz}
\author{Sun-Ju Chung}
\affiliation{Korea Astronomy and Space Science Institute, Daejon 34055, Republic of Korea}
\email{sjchung@kasi.re.kr}
\author{Andrew Gould}
\affiliation{Department of Astronomy, Ohio State University, 140 West 18th Ave., Columbus, OH 43210, USA}
\email{gould.34@osu.edu}
\author{Youn Kil Jung}
\affiliation{Korea Astronomy and Space Science Institute, Daejon 34055, Republic of Korea}
\affiliation{University of Science and Technology, Daejeon 34113, Republic of Korea}
\email{younkil21@gmail.com}
\author{Kyu-Ha Hwang}
\affiliation{Korea Astronomy and Space Science Institute, Daejon 34055, Republic of Korea}
\email{kyuha@kasi.re.kr}
\author{Yoon-Hyun Ryu}
\affiliation{Korea Astronomy and Space Science Institute, Daejon 34055, Republic of Korea}
\email{yhryu@kasi.re.kr}
\author{Yossi Shvartzvald}
\affiliation{Department of Particle Physics and Astrophysics, Weizmann Institute of Science, Rehovot 76100, Israel}
\email{yossishv@gmail.com}
\author{In-Gu Shin}
\affiliation{Department of Astronomy, Westlake University, Hangzhou 310030, Zhejiang Province, China}
\email{ingushin@gmail.com}
\author{Jennifer C. Yee}
\affiliation{Center for Astrophysics $|$ Harvard \& Smithsonian 60 Garden St., Cambridge, MA 02138, USA}
\email{jyee@cfa.harvard.edu}
\author{Hongjing Yang}
\affiliation{Department of Astronomy, Westlake University, Hangzhou 310030, Zhejiang Province, China}
\email{yanghongjing@westlake.edu.cn}
\author{Doeon Kim}
\affiliation{Department of Physics, Chungbuk National University, Cheongju 28644, Republic of Korea}
\email{qso21@hanmail.net}
\author{Dong-Jin Kim}
\affiliation{Korea Astronomy and Space Science Institute, Daejon 34055, Republic of Korea}
\email{keaton03@kasi.re.kr}
\author{Seung-Lee Kim}
\affiliation{Korea Astronomy and Space Science Institute, Daejon 34055, Republic of Korea}
\email{slkim@kasi.re.kr}
\author{Dong-Joo Lee}
\affiliation{Korea Astronomy and Space Science Institute, Daejon 34055, Republic of Korea}
\email{marin678@kasi.re.kr}
\author{Sang-Mok Cha}
\affiliation{Korea Astronomy and Space Science Institute, Daejon 34055, Republic of Korea}
\email{chasm@kasi.re.kr}
\author{Yongseok Lee}
\affiliation{Korea Astronomy and Space Science Institute, Daejon 34055, Republic of Korea}
\email{yslee@kasi.re.kr}
\author{Byeong-Gon Park}
\affiliation{Korea Astronomy and Space Science Institute, Daejon 34055, Republic of Korea}
\email{bgpark@kasi.re.kr}
\author{Richard W. Pogge}
\affiliation{Department of Astronomy, Ohio State University, 140 West 18th Ave., Columbus, OH 43210, USA}
\email{pogge.1@osu.edu}
\author{Hongyu Li}
\affiliation{Department of Astronomy, Tsinghua University, Beijing 100084, China}
\email{lihongyu25@mails.tsinghua.edu.cn}
\author[0000-0002-1279-0666]{Jiyuan Zhang}
\affiliation{Department of Astronomy, Tsinghua University, Beijing 100084, China}
\email{zhangjy22@mails.tsinghua.edu.cn}
\author[0000-0001-8317-2788]{Shude Mao}
\affiliation{Department of Astronomy, Westlake University, Hangzhou 310030, Zhejiang Province, China}
\email{shude.mao@westlake.edu.cn}
\collaboration{30}{(KMTNet Collaboration)}
\author{Przemek Mr{\'o}z}
\affiliation{Astronomical Observatory, University of Warsaw, Al.~Ujazdowskie 4, 00-478 Warszawa, Poland}
\email{pmroz@astrouw.edu.pl}
\author{Micha{\l} K. Szyma{\'n}ski}
\affiliation{Astronomical Observatory, University of Warsaw, Al.~Ujazdowskie 4, 00-478 Warszawa, Poland}
\email{msz@astrouw.edu.pl}
\author{Jan Skowron}
\affiliation{Astronomical Observatory, University of Warsaw, Al.~Ujazdowskie 4, 00-478 Warszawa, Poland}
\email{jskowron@astrouw.edu.pl}
\author{Rados{\l}aw Poleski} 
\affiliation{Astronomical Observatory, University of Warsaw, Al.~Ujazdowskie 4, 00-478 Warszawa, Poland}
\email{radek.poleski@gmail.co}
\author{Igor Soszy{\'n}ski}
\affiliation{Astronomical Observatory, University of Warsaw, Al.~Ujazdowskie 4, 00-478 Warszawa, Poland}
\email{soszynsk@astrouw.edu.pl}
\author{Pawe{\l} Pietrukowicz}
\affiliation{Astronomical Observatory, University of Warsaw, Al.~Ujazdowskie 4, 00-478 Warszawa, Poland}
\email{pietruk@astrouw.edu.pl}
\author{Szymon Koz{\l}owski} 
\affiliation{Astronomical Observatory, University of Warsaw, Al.~Ujazdowskie 4, 00-478 Warszawa, Poland}
\email{simkoz@astrouw.edu.pl}
\author{Krzysztof A. Rybicki}
\affiliation{Astronomical Observatory, University of Warsaw, Al.~Ujazdowskie 4, 00-478 Warszawa, Poland}
\email{krybicki@astrouw.edu.pl}
\author{Patryk Iwanek}
\affiliation{Astronomical Observatory, University of Warsaw, Al.~Ujazdowskie 4, 00-478 Warszawa, Poland}
\email{piwanek@astrouw.edu.pl}
\author{Krzysztof Ulaczyk}
\affiliation{Department of Physics, University of Warwick, Gibbet Hill Road, Coventry, CV4 7AL, UK}
\email{kulaczyk@astrouw.edu.pl}
\author{Marcin Wrona}
\affiliation{Astronomical Observatory, University of Warsaw, Al.~Ujazdowskie 4, 00-478 Warszawa, Poland}
\affiliation{Villanova University, Department of Astrophysics and Planetary Sciences, 800 Lancaster Ave., Villanova, PA 19085, USA}
\email{mwrona@astrouw.edu.pl}
\author{Mariusz Gromadzki}          
\affiliation{Astronomical Observatory, University of Warsaw, Al.~Ujazdowskie 4, 00-478 Warszawa, Poland}
\email{marg@astrouw.edu.pl}
\author{Mateusz J. Mr{\'o}z} 
\affiliation{Astronomical Observatory, University of Warsaw, Al.~Ujazdowskie 4, 00-478 Warszawa, Poland}
\email{mmroz@astrouw.edu.pl}
\collaboration{100}{(The OGLE Collaboration)}
\correspondingauthor{\texttt{cheongho@astroph.chungbuk.ac.kr}}
\correspondingauthor{\texttt{leecu@kasi.re.kr}}

\begin{abstract}
We analyze the microlensing event OGLE-2018-BLG-0829, whose light curve exhibits 
a weak short-term anomaly in addition to the prominent caustic-crossing features 
of a binary lens. The anomaly cannot be explained by a binary-lens single-source 
(2L1S) model.  We therefore investigate both binary-lens binary-source (2L2S) and 
triple-lens single-source (3L1S) interpretations and find that both models reproduce 
the additional anomaly, with the 3L1S model being favored by $\Delta\chi^2=15.1$.  
Under the favored 3L1S interpretation, the anomaly arises from a caustic induced 
by a planetary-mass companion, whereas the prominent caustic-crossing features are 
produced by the underlying binary lens.  Combining the measured event timescale, 
$t_{\rm E}=7.55\pm0.30$ days, and angular Einstein radius, $\theta_{\rm E}=0.325 
\pm 0.059$ mas, with a Bayesian analysis, we infer a lens system consisting of an 
M dwarf of mass $0.193^{+0.300}_{-0.098}\,M_\odot$, a possible brown-dwarf companion 
of mass $0.068^{+0.105}_{-0.034}\,M_\odot$, and a Saturn-mass planet of mass 
$0.337^{+0.522}_{-0.171}\,M_{\rm J}$. The companions have projected separations 
of approximately 3.6 au and 1.6 au from the primary, respectively.  However, the 
physical interpretation remains tentative because potentially important higher-order 
effects have not been considered, and the dynamical stability of the proposed 
configuration remains uncertain. If confirmed, this would be the third star--brown 
dwarf--planet system identified through microlensing, highlighting the potential 
of microlensing to detect planets in complex multiple systems.
\end{abstract}


\section{Introduction \label{sec:one}} 

Most stars are thought to form in binary or higher-order multiple systems rather than 
in isolation \citep{Duquennoy1991,Raghavan2010}. Consequently, understanding planet
formation requires investigating not only planets orbiting single stars, but also those
residing in dynamically complex multiple systems. The presence of a stellar or substellar
companion can substantially modify the structure and evolution of a protoplanetary disk
through tidal truncation, enhanced dynamical perturbations, and accelerated disk
dispersal, thereby influencing both the formation and long-term survival of planets
\citep{Artymowicz1994,Kraus2012}. Planetary systems in multiple-star environments
therefore provide important observational constraints on theories of planet formation
under conditions that differ markedly from those around isolated stars.

Brown dwarfs occupy the mass range between stars and planets and are thought to form
through a variety of mechanisms, including turbulent fragmentation, disk fragmentation,
and dynamical ejection \citep{Whitworth2007,Luhman2012}. Their role in planetary
systems remains poorly understood. In particular, whether planets can efficiently form
and survive around stars possessing brown-dwarf companions remains an open question.
Such systems probe a poorly explored regime in which the gravitational influence of the
brown dwarf may significantly alter the conditions for planet formation and subsequent
planetary evolution.

Gravitational microlensing is uniquely sensitive to planets located beyond the snow line
around low-mass stars and is largely independent of the luminosities of either the host
or the planet \citep{Mao1991,Gould1992b,Gaudi2012}. Unlike other detection methods,
microlensing can detect planets orbiting faint M dwarfs, brown dwarfs, stellar remnants,
and even free-floating objects \citep[e.g.,][]{Bond2004,Beaulieu2006,Gaudi2008,
Bennett2014,Mroz2017}. It is also capable of revealing planets in binary and higher-order
multiple systems through the characteristic perturbations they produce in lensing caustics
\citep{Lee2008,Han2017}.

To date, only two microlensing planetary systems containing both a stellar host 
and a brown-dwarf companion have been reported. The first, KMT-2020-BLG-0414Lb 
\citep{Zang2021}, consists of an M dwarf hosting an Earth-mass planet together 
with a brown-dwarf companion. More recently, \citet{Han2025} reported KMT-2024-BLG-0404L, 
in which a Uranus-mass planet orbits a close binary composed of a late M dwarf and 
a brown dwarf. Although these discoveries demonstrate that planets can exist in such 
environments, the sample remains extremely limited. Each additional system therefore 
provides valuable constraints on planet formation and dynamical evolution in the 
presence of brown-dwarf companions.

Interpreting multiple-lens microlensing events, however, is challenging because 
their anomalies can exhibit a wide variety of morphologies arising from the 
interplay of multiple caustics \citep{Danek2015,Danek2019,Bozza2000,Han2001}.  
This diversity frequently gives rise to degenerate or non-unique solutions, making 
detailed light-curve modeling essential for identifying the correct lens configuration. 
In some cases, the presence of an additional planet is revealed only through a subtle 
perturbation to the caustic structure generated by the underlying binary lens, making 
its detection considerably more difficult than in ordinary planetary microlensing 
events. Expanding the sample of such systems is therefore important not only for 
understanding planet formation in dynamically perturbed environments but also for 
exploring the full discovery space of gravitational microlensing.

In this paper, we present an analysis of the microlensing event OGLE-2018-BLG-0829, 
whose light curve exhibits a weak anomaly in addition to the prominent caustic-crossing 
features of a binary lens.  We investigate two possible explanations for the weak 
anomaly: a binary-source interpretation and a triple-lens interpretation. Our analysis 
favors the triple-lens model, in which the lens system consists of a stellar primary 
and two lower-mass companions. According to the favored model, Bayesian analysis 
suggests that the system comprises an M dwarf, a brown dwarf, and a Saturn-mass 
planet, although the inferred masses are subject to substantial uncertainties. If 
confirmed, this would be the third star--brown-dwarf--planet system discovered by 
microlensing.  This event illustrates the potential of microlensing to detect planets 
through subtle perturbations in binary-lens light curves, while highlighting the 
challenges of uniquely characterizing such complex lens systems.

\begin{figure*}[t]
\centering
\includegraphics[width=15.0cm]{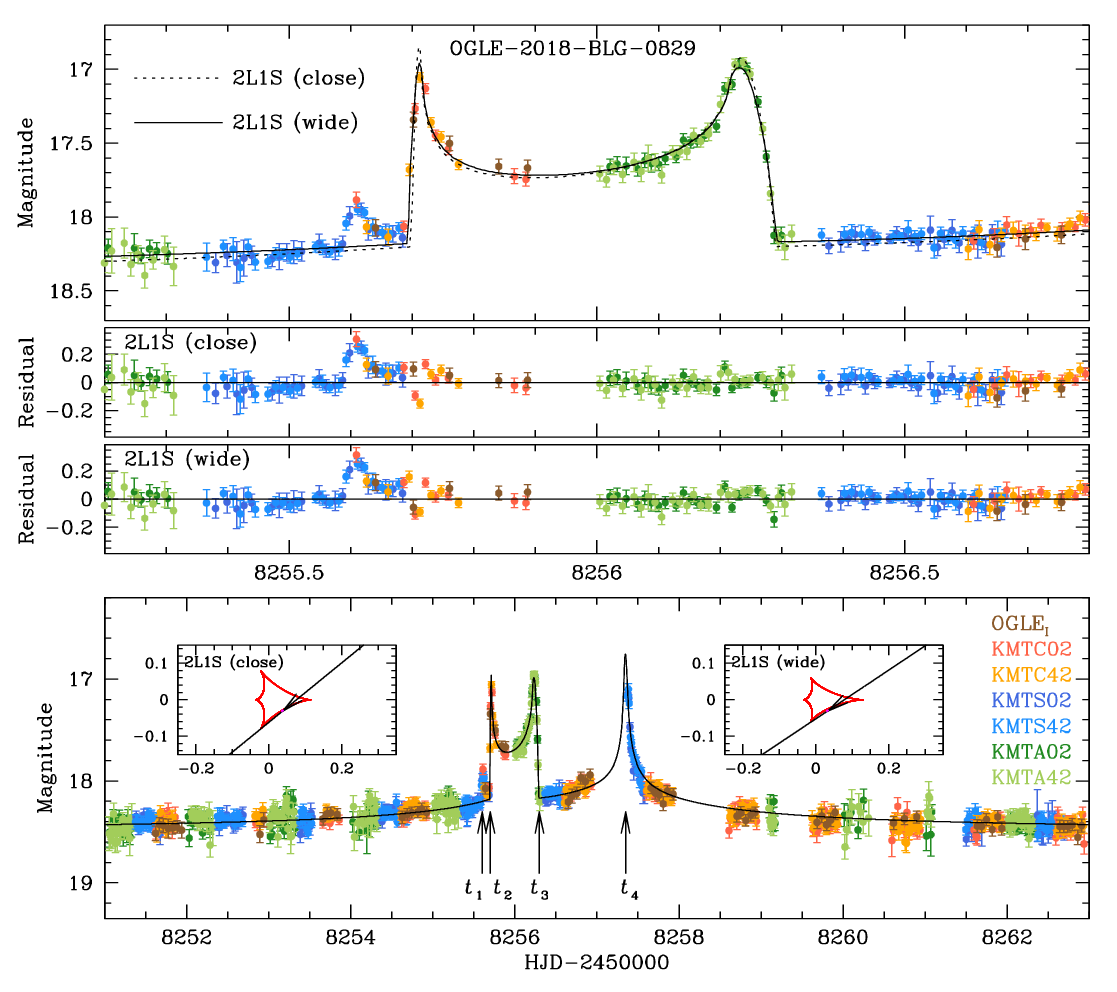}
\caption{
Light curve of OGLE-2018-BLG-0829.  The bottom panel shows the full light curve, 
while the upper panels present an enlarged view of the anomalous region around 
the caustic-crossing features and the residuals from the close and wide 2L1S 
models.  The dotted and solid curves represent the close and wide 2L1S solutions, 
respectively.  The arrows labeled $t_1$, $t_2$, $t_3$, and $t_4$ indicate the 
four anomaly features discussed in the text.  The two insets in the bottom panel 
illustrate the lens configurations of the close and wide 2L1S models.  In each 
inset, the cuspy closed curve represents the caustic, and the arrowed line denotes 
the source trajectory.
}
\label{fig:one}
\end{figure*}

\section{Observations and data \label{sec:two}}

The source star of the microlensing event OGLE-2018-BLG-0829 is located at 
equatorial coordinates $(\alpha,\delta)_{\rm J2000}=(17{:}55{:}19.74, -29{:}47{:}08.9)$, 
corresponding to Galactic coordinates $(l, b)=(0\fdg 3539,-2\fdg 2520)$.  The event 
lies in a field with an $I$-band extinction of $A_I=1.68$ \citep{Nataf2013} and has 
a baseline magnitude of $I_{\rm base}=18.53$.

The event was discovered by the Optical Gravitational Lensing Experiment (OGLE) 
survey \citep{Udalski2015}. The OGLE survey conducts wide-field photometric 
monitoring of the Galactic bulge stars using the 1.3~m Warsaw Telescope at Las 
Campanas Observatory in Chile.  Although  the Korea Microlensing Telescope Network 
(KMTNet; \citealt{Kim2016}) group did not identify or issue an alert for the event 
during the 2018 observing season, the event was located in the region covered by the 
overlapping KMTNet prime fields BLG02 and BLG03.  This enabled us to recover the light 
curve by performing post-season photometric reductions of the archived KMTNet images.  
KMTNet consists of three identical 1.6~m telescopes, each equipped with a 
$4~\mathrm{deg}^2$ camera, located at the Cerro Tololo Inter-American Observatory 
in Chile (KMTC), the South African Astronomical Observatory in South Africa (KMTS), 
and the Siding Spring Observatory in Australia (KMTA).  This global network provides 
nearly continuous photometric coverage of lensing events toward the Galactic bulge.

\begin{figure*}[t]
\centering
\includegraphics[width=15.0cm]{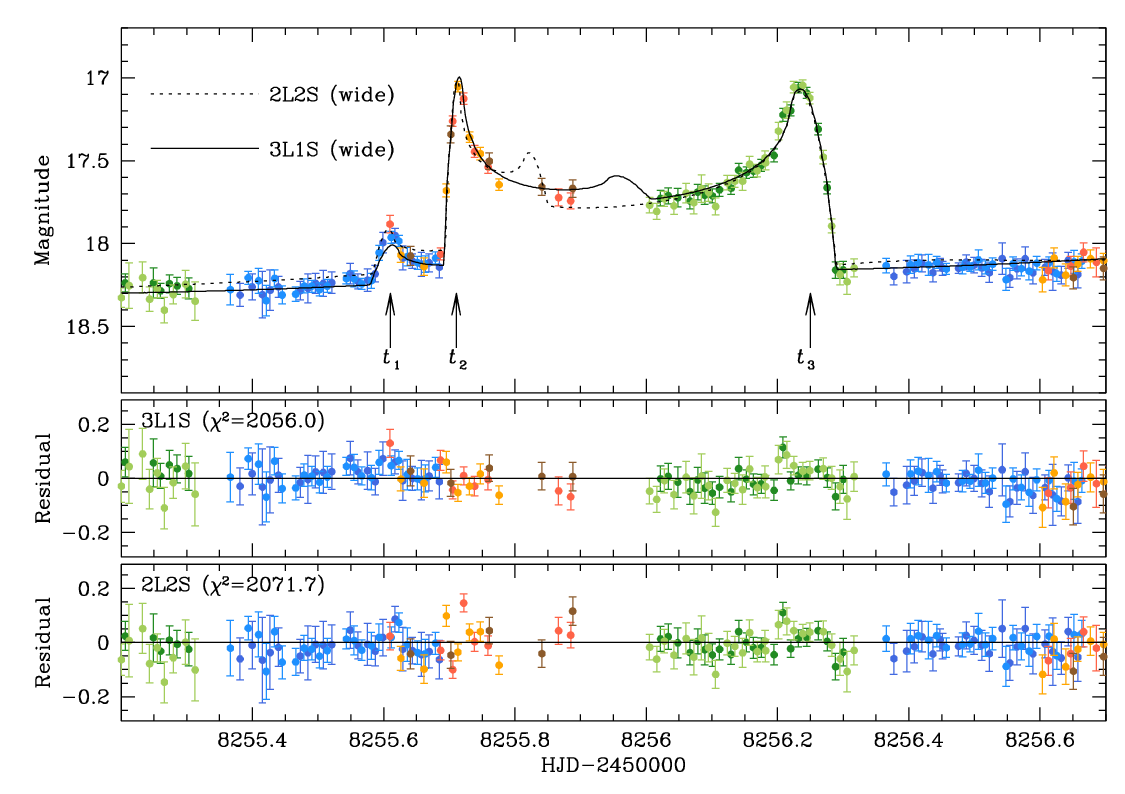}
\caption{
Enlarged view of the anomaly region around the caustic-crossing features with the 
best-fit 2L2S and 3L1S models overlaid.  The labels $t_1$--$t_3$ correspond to the 
anomaly features marked in Fig.~\ref{fig:one}.
}
\label{fig:two}
\end{figure*}

Our initial analysis was based on the OGLE photometry together with the standard 
KMTNet pipeline photometry. These data clearly revealed the major caustic-crossing 
features characteristic of a binary-lens event, leading us to initially interpret 
the event as a conventional binary-lens event.  However, the relatively large 
photometric scatter in the pipeline reduction obscured weaker deviations from the 
binary-lens light curve, preventing us from recognizing the presence of an additional 
anomaly.

The subtle anomaly became apparent only after the KMTNet images were reprocessed 
using the pySIS photometric package developed by \citet{Yang2024} for high-precision 
photometry of KMTNet microlensing events. The resulting light curve is presented in 
Figure~\ref{fig:one}.  At first glance, the light curve is dominated by the characteristic 
signatures of a binary-lens caustic crossing: a pair of sharp caustic-crossing spikes 
at $t_2$ (${\rm HJD}^\prime \equiv {\rm HJD} - 2450000 \sim 8255.7$) and $t_3$ 
(${\rm HJD}^\prime \sim 8256.3$), followed by a broad peak at $t_4$ (${\rm HJD}^\prime 
\sim 8257.3$) produced by the source's close approach to a caustic cusp. In addition 
to these prominent features, however, the improved photometry reveals a weak but 
statistically significant bump at $t_1$ (${\rm HJD}^\prime \sim 8255.6$), occurring 
shortly before the first caustic-crossing feature.  This previously unrecognized 
feature motivates our investigation of more complex lensing configurations to determine 
its possible origin.

\section{Binary-lens single-source (2L1S) interpretation \label{sec:three}}

The prominent caustic-crossing features indicate that the lens system contains 
multiple masses.  We therefore begin by modeling the light curve with a binary-lens 
single-source (2L1S) configuration, the simplest lens model capable of producing 
caustics. The modeling is performed by searching for the set of lensing parameters 
that most accurately reproduces the observed light curve.

For a 2L1S event, the light curve is described by seven principal parameters.  Three 
of these characterize the source trajectory relative to the lens, $(t_0, u_0, \te)$, 
where $t_0$ is the time of closest approach between the source and the lens, $u_0$ 
is the impact parameter in units of the angular Einstein radius, and $\te$ is the 
Einstein timescale. Two parameters, $(s, q)$, describe the binary lens, where $s$ 
is the projected separation between the lens components normalized to the angular 
Einstein radius and $q$ is their mass ratio.  The remaining two parameters are the 
angle of the source trajectory with respect to the binary axis, $\alpha$, and the 
normalized source radius, $\rho=\theta_*/\thetae$, where $\theta_*$ and $\thetae$ 
denote the angular source radius and angular Einstein radius, respectively. 
The normalized source radius is included to account for finite-source effects during 
caustic crossings, which smooth the sharp magnification variations predicted for a 
point source because different parts of the extended source experience different 
magnifications \citep{Schneider1987,Gould1994, Dominik2004}.

\begin{deluxetable*}{l|cc|cccc}
\tablewidth{0pt}
\tablecaption{Best-fit lensing parameters for the 2L2S and 3L1S models. \label{table:one}}
\tablehead{
\multicolumn{1}{c|}{Parameter}        &
\multicolumn{2}{c|}{2L2S}             &
\multicolumn{2}{c}{3L1S}             \\
\multicolumn{1}{c|}{}                 &
\multicolumn{1}{c}{Close}            &
\multicolumn{1}{c|}{Wide }            &
\multicolumn{1}{c}{Close}            &
\multicolumn{1}{c}{Wide }   
}
\startdata
 $\chi^2$                     &  $2107.1            $  &  $2071.1            $   &  $2147.7            $   &  $2056.0            $ \\ 
 $t_0$ (HJD$^\prime$)         &  $8256.540 \pm 0.015$  &  $8256.780 \pm 0.017$   &  $8256.332 \pm 0.013$   &  $8256.650 \pm 0.020$ \\
 $u_0$                        &  $0.0443 \pm  0.0015$  &  $0.0544 \pm 0.0012 $   &  $0.0796 \pm 0.0011 $   &  $0.0761 \pm 0.0027 $ \\
 $\te$ (days)                 &  $13.16 \pm 0.48    $  &  $10.46 \pm 0.33    $   &  $6.57 \pm 0.10     $   &  $7.55 \pm 0.30     $ \\ 
 $s_2$                        &  $0.536 \pm 0.012   $  &  $2.156 \pm 0.052   $   &  $0.6200 \pm 0.0034 $   &  $2.160 \pm 0.043   $ \\
 $q_2$                        &  $0.1217 \pm 0.0091 $  &  $0.195 \pm 0.017   $   &  $0.204 \pm 0.009   $   &  $0.338 \pm 0.019   $ \\ 
 $\alpha$ (rad)               &  $5.591 \pm 0.016   $  &  $5.730 \pm 0.015   $   &  $5.458 \pm 0.013   $   &  $5.624 \pm 0.016   $ \\
 $s_3$                        &   \nodata              &   \nodata               &  $1.1184 \pm 0.0020 $   &  $0.9328 \pm 0.0087 $ \\ 
 $q_3$ ($10^{-3}$)            &   \nodata              &   \nodata               &  $0.675 \pm 0.079   $   &  $1.57 \pm 0.18     $ \\ 
 $\psi$ (rad)                 &   \nodata              &   \nodata               &  $0.660 \pm 0.014   $   &  $0.725 \pm 0.018   $ \\ 
 $t_{0,2}$ (HJD$^\prime$)     &  $8256.040 \pm 0.016$  &  $8255.939 \pm 0.012$   &   \nodata               &   \nodata             \\    
 $u_{0,2}$                    &  $-0.0365 \pm 0.0014$  &  $-0.0390 \pm 0.0010$   &   \nodata               &   \nodata             \\    
 $\rho_1$ ($10^{-3}$)         &  $0.53 \pm 0.02     $  &  $0.77 \pm 0.03     $   &  $1.214 \pm 0.059   $   &  $1.534 \pm 0.071   $ \\  
 $\rho_2$ ($10^{-3}$)         &  $1.34 \pm 0.31     $  &  $1.40 \pm 0.32     $   &   \nodata               &   \nodata             \\
 $q_F$                        &  $0.196 \pm 0.015   $  &  $0.166 \pm 0.012   $   &   \nodata               &   \nodata             \\
\enddata
\tablecomments{
 ${\rm HJD}^\prime\equiv {\rm HJD}-2450000$.
}
\end{deluxetable*}

A standard 2L1S analysis yields a pair of local solutions arising from the well-known 
close--wide degeneracy \citep{Dominik1999, An2005}.  This degeneracy occurs when binary 
lenses with separations smaller and larger than the angular Einstein radius produce 
similar central caustic structures, resulting in light curves with similar anomaly 
patterns. In this case, the close solution has lensing parameters $(s, q, \te) \sim 
(0.55, 0.12, 11.9\,{\rm days})$, while the wide solution has $(2.32, 0.19, 
13.4\,{\rm days})$. Although both solutions reproduce the major features of the 
observed light curve, the wide solution provides a better fit, with $\Delta\chi^2 
\sim 46.4$. The model light curves of the two solutions are overlaid on the data 
in Figure~\ref{fig:one}.

The lens-system configurations corresponding to the close and wide 2L1S solutions are 
shown in the two insets of the bottom panel of Figure~\ref{fig:one}. In both cases, the 
source traverses the right side of the central caustic associated with the primary lens 
component ($M_1$). The caustic entrance and exit produce the sharp spikes at $t_2$ and 
$t_3$, respectively, while the subsequent close approach to the lower off-axis caustic 
cusp gives rise to the broad peak at $t_4$.

Although both solutions adequately reproduce the three major anomaly features at $t_2$, 
$t_3$, and $t_4$, neither accounts for the weak bump at $t_1$. The persistence of this 
residual indicates that the anomaly cannot be explained by a simple binary-lens geometry 
alone and instead requires an additional lens or source component. Motivated by this 
discrepancy, we explore more complex lensing configurations in the following sections 
to investigate the possible origin of the $t_1$ anomaly and identify plausible 
interpretations of the event.

\section{Binary-lens binary-source (2L2S) interpretation \label{sec:four}}

We first tested a binary-lens binary-source (2L2S) model, motivated by the similarity 
of the observed anomaly to those in the microlensing events OGLE-2016-BLG-1003 
\citep{Jung2017} and KMT-2021-BLG-0209 \citep{Han2026}. In both events, the 
short-term anomaly was interpreted as resulting from the caustic crossing of a 
secondary source star located close to the primary source.

Compared to the 2L1S configuration, the 2L2S model introduces an additional source 
star, $S_2$.  Modeling the anomaly produced by $S_2$ requires four additional lensing 
parameters, $(t_{0,2}, u_{0,2}, \rho_2, q_{\rm F})$.  The parameters $t_{0,2}$ and 
$u_{0,2}$ represent the time at which $S_2$ approaches closest to the reference 
position of the lens and the corresponding lens--source separation, respectively.  
The third parameter, $\rho_2$, is the normalized source radius of $S_2$, while the 
fourth parameter, $q_{\rm F}$, represents the flux ratio between the secondary and 
primary ($S_1$) source stars.

The 2L2S modeling was initialized with the best-fit 2L1S solution, which successfully 
reproduces the overall light curve except for the anomaly at $t_1$. We then searched 
for the trajectory of the secondary source that best accounts for the timing and 
amplitude of the unexplained anomaly at $t_1$.

\begin{figure}[t]
\includegraphics[width=\columnwidth]{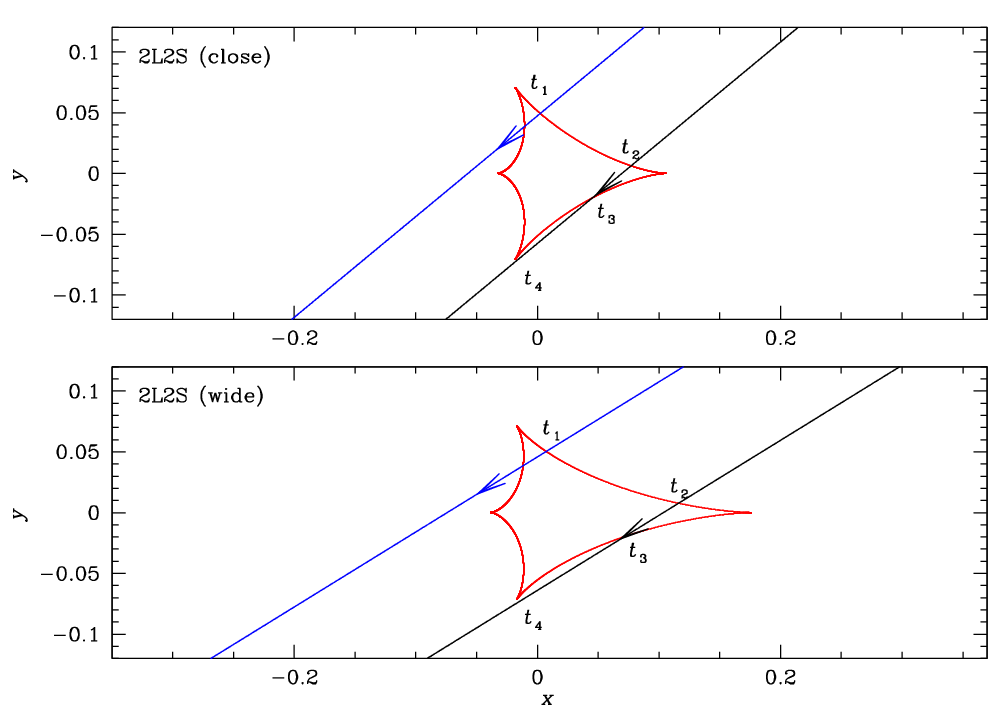}
\caption{
Lens-system configurations of the close and wide 2L2S solutions. The black and 
blue arrowed lines represent the trajectories of the primary and secondary source 
stars, respectively.  The points labeled $t_1$, $t_2$, $t_3$, and $t_4$ mark the 
source positions corresponding to the anomaly features indicated in Figure~\ref{fig:one}. 
In both panels, the secondary source crosses the upper cusp of the caustic, producing 
the anomaly at $t_1$.
}
\label{fig:three}
\end{figure}

The modeling yielded a pair of local solutions originating from the close and 
wide 2L1S baseline models. We designate these as the ``close'' and ``wide'' 2L2S 
solutions, respectively. The lensing parameters of the two solutions are listed 
in Table~\ref{table:one} together with their $\chi^2$ values. We find that the 
wide solution is favored over the close solution by $\Delta\chi^2 = 36.0$.  The 
model light curve of the preferred wide solution is shown as the dotted curve in 
Figure~\ref{fig:two}.  The model approximately reproduces all the observed anomaly 
features, including the previously unexplained bump at $t_1$.

The lens-system configurations of the two 2L2S solutions are presented in 
Figure~\ref{fig:three}.  The caustic geometries are nearly identical to those of 
the corresponding 2L1S solutions, and the trajectory of the primary source relative 
to the caustic remains essentially unchanged. The principal difference is the presence 
of the secondary source, which passes over the upper cusp of the caustic.  The secondary 
source is relatively faint, with a flux ratio of $q_{\rm F} \sim 0.17$ relative to the 
primary source. It leads the primary source by $\Delta t \sim0.84$\,days, entering the 
caustic ahead of the primary and thereby producing the anomaly at $t_1$. According to 
the model, the secondary source exits the caustic at around ${\rm HJD}^\prime \sim 
8255.8$, while the primary source is traversing the interior of the caustic. This 
caustic exit of $S_2$ produces a weak caustic-crossing feature, but the feature is 
only sparsely sampled by the observations.

\begin{figure}[t]
\includegraphics[width=\columnwidth]{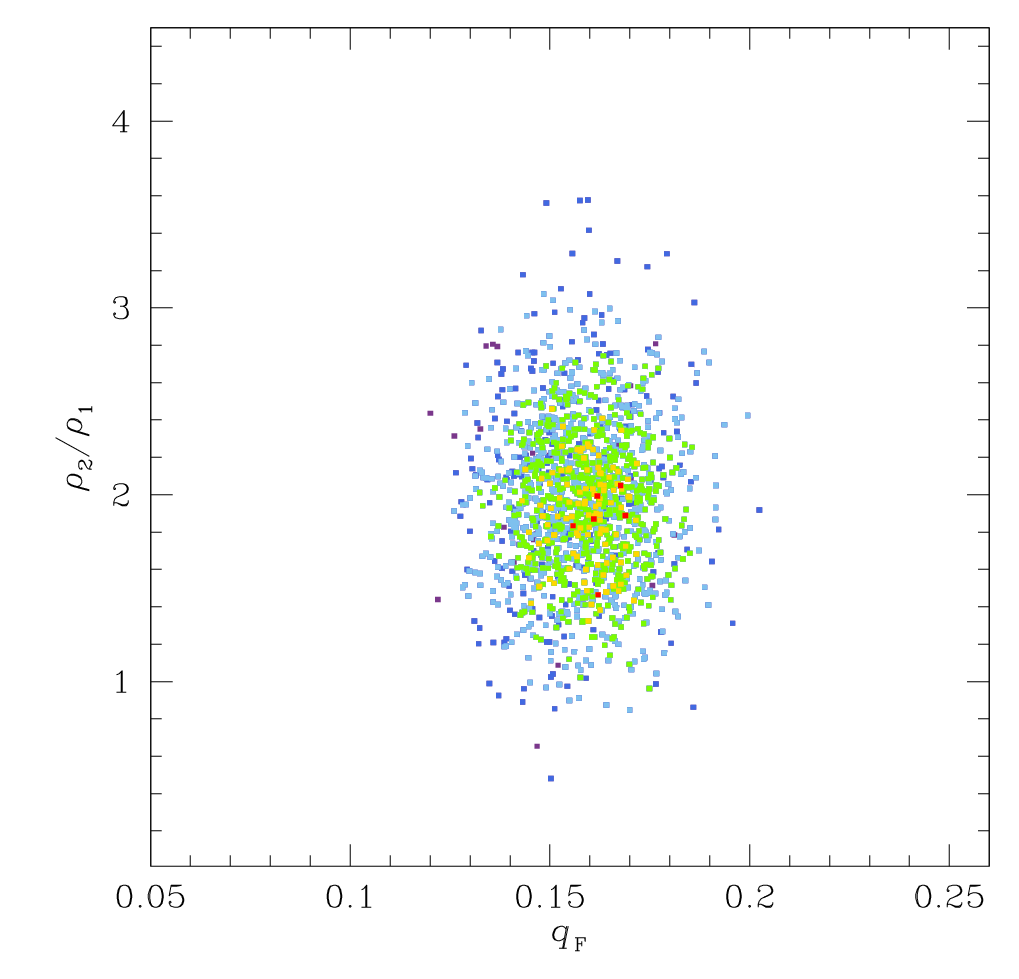}
\caption{
Scatter plot of the MCMC samples on the $\rho_2/\rho_1$--$q_{\rm F}$ plane for 
the wide 2L2S solution.  The points are color-coded according to their confidence 
levels: $<1\sigma$ (red), $1$--$2\sigma$ (yellow), $2$--$3\sigma$ (green), 
$3$--$4\sigma$ (cyan), and $4$--$5\sigma$ (blue). 
}
\label{fig:four}
\end{figure}

\begin{figure*}[t]
\centering
\includegraphics[width=15.0cm]{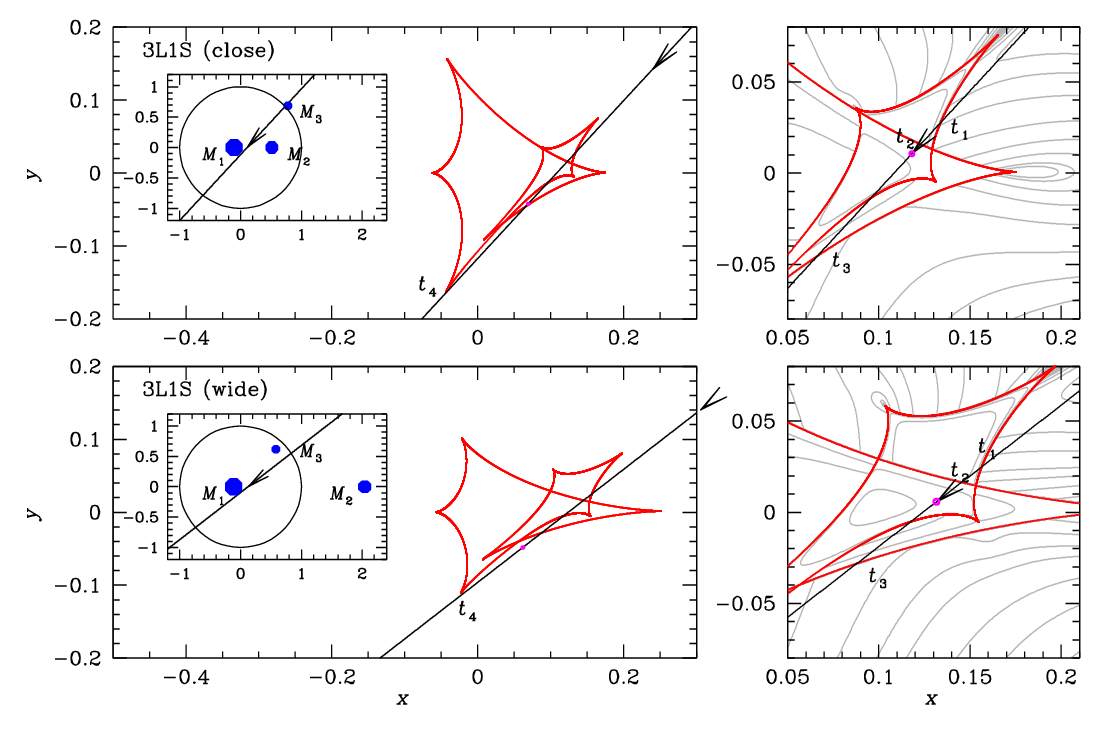}
\caption{
Lens-system configurations of the close (upper panels) and wide (lower panels) 
3L1S solutions.  For each solution, the left panel shows the full caustic 
structure, while the right panel presents an enlarged view of the source-crossing 
region.  The inset in each left panel shows the lens components as filled blue 
circles and the Einstein ring as a solid circle of unit radius.  In the right 
panels, the points labeled $t_1$, $t_2$, $t_3$, and $t_4$ along the source 
trajectory indicate the source positions corresponding to the anomaly features 
marked in Fig.~\ref{fig:one}.  The gray curves surrounding the caustics are 
equal-magnification contours.
}
\label{fig:five}
\end{figure*}

Although the 2L2S model provides a reasonable fit to the data, several considerations 
argue against this interpretation.  First, the inferred source properties appear 
difficult to reconcile with those expected for a typical binary-source system. The 
normalized source radii of the primary and secondary source stars inferred from 
the 2L2S model are $\rho_1 = (0.77 \pm 0.03)\times 10^{-3}$ and $\rho_2 = (1.40 
\pm 0.32)\times 10^{-3}$, respectively, implying that $S_2$ is larger than $S_1$. 
However, the measured flux ratio is only $q_{\rm F} = 0.166 \pm 0.012$, indicating 
that $S_2$ is substantially fainter than $S_1$. This relationship is illustrated in 
Figure~\ref{fig:four}, which shows the distribution of the Markov Chain Monte Carlo 
(MCMC) samples on the $\rho_2/\rho_1$--$q_{\rm F}$ plane. The MCMC samples favor 
$\rho_2/\rho_1 > 1$ despite the small flux ratio.  Although such a configuration 
is possible if the larger source has a sufficiently lower effective temperature, 
it would require an unusual combination of stellar properties for two stars with 
similar ages and metallicities, as expected for a physically bound binary. We 
therefore regard this configuration as less plausible, although it cannot be ruled 
out solely on the basis of the inferred source properties.  Second, the 2L2S model 
is statistically disfavored relative to the 3L1S model discussed in the following 
subsection, providing a poorer fit by $\Delta\chi^2 = 15.1$.

\section{Triple-lens single-source (3L1S) interpretation \label{sec:five}}

We also tested a triple-lens single-source (3L1S) model, in which the lens 
consists of three masses and the source is a single star. Compared to the 2L1S 
configuration, the 3L1S model introduces an additional lens component and 
therefore requires three extra lensing parameters, $(s_3, q_3, \psi)$. Here, 
$s_3$ is the projected separation between the third lens component ($M_3$) and 
the primary lens ($M_1$), normalized to the angular Einstein radius. The parameter 
$q_3$ denotes the mass ratio of $M_3$ to $M_1$, and $\psi$ is the position angle 
of $M_3$ measured with respect to the primary--secondary binary-lens axis. To 
distinguish these parameters from those of the first companion ($M_2$), we denote 
the separation and mass ratio of the latter by $(s_2, q_2)$.

To characterize the third lens component, we first conducted a grid search over 
the $(s_3, q_3, \psi)$ parameter space. The remaining lensing parameters were 
initially fixed at their best-fit 2L1S values, because the 2L1S model adequately 
reproduces the overall light-curve morphology except for the anomaly at $t_1$. We 
then constructed a $\chi^2$ map over the grid parameter space to identify local 
minima. Each identified minimum was subsequently refined by allowing all lensing 
parameters to vary.

This procedure yielded a pair of local solutions originating from the close and 
wide 2L1S solutions.  The lensing parameters of the two 3L1S solutions are listed 
in Table~\ref{table:one}. The wide solution is favored over the close solution by 
$\Delta\chi^2 = 91.7$, a sufficiently large difference to resolve the close--wide 
degeneracy. The model light curve of the preferred wide solution, together with 
the corresponding residuals, is shown in Figure~\ref{fig:two}.

For the preferred wide solution, the triple-lens parameters are
$(s_2, q_2, s_3, q_3) \sim (2.16, 0.34, 0.93, 1.6 \times 10^{-3})$.
These values suggest that the lens is a triple system consisting of a binary 
with a projected separation of approximately twice the angular Einstein radius 
and a low-mass third companion located at a projected separation close to the 
angular Einstein radius from the primary. The small mass ratio of the third 
component, $q_3 \sim 1.6 \times 10^{-3}$, suggests its planetary nature. For 
this mass ratio, the companion mass would fall below the conventional 
planet--brown-dwarf boundary of $\sim 13\,M_{\rm J}$ for a primary mass below 
$\sim 7.8\,M_\odot$. As discussed in Section~\ref{sec:seven}, the Bayesian 
analysis favors a low-mass stellar primary, supporting the planetary interpretation.
The Einstein timescale of the event, $\te \sim 7.6$~days, is relatively short, 
which may indicate a low-mass lens system. However, a short timescale can also 
result from a small lens--source relative parallax or a high relative proper 
motion, and therefore does not necessarily imply low lens masses.

The lens-system configurations of the close and wide 3L1S solutions are 
presented in Figure~\ref{fig:five}.  In both solutions, the planet generates 
an additional caustic above the central caustic produced by the $M_1$--$M_2$ 
binary.  Consistent with the binary-superposition approximation proposed by 
\citet{Bozza2000} and \citet{Han2001}, the overall caustic topology is well 
described by the superposition of the binary caustics associated with the 
$M_1$--$M_2$ and $M_1$--$M_3$ lens pairs.

According to the preferred model, the source first enters the planet-induced 
caustic, producing the anomaly at $t_1$, and subsequently enters the binary-induced 
caustic, generating the caustic spike at $t_2$. It then exits the planet-induced 
caustic before leaving the binary caustic, producing a weak caustic-crossing 
feature at around ${\rm HJD}^\prime \sim 8255.9$ within the U-shaped trough. 
This predicted feature was not covered by the observations. The source subsequently 
exits the binary caustic at $t_3$, producing the second caustic-crossing spike, and 
finally passes close to the lower cusp of the binary caustic after leaving the 
caustic, generating the broad peak at $t_4$.

\section{Source star and angular Einstein radius \label{sec:six}}

Characterizing the source star is an essential step in measuring the angular Einstein 
radius, $\thetae$. Once the source type is determined, its angular radius, $\theta_*$, 
can be estimated from its de-reddened color and magnitude. The angular Einstein radius 
is then obtained from
\begin{equation}
\thetae = \frac{\theta_*}{\rho},
\label{eq1}
\end{equation}
\hskip-4pt
where $\rho$ is the normalized source radius measured from the light-curve modeling. 
The angular Einstein radius provides an important constraint on the physical properties 
of the lens because it is related to the lens mass and distance through
\begin{equation}
\thetae = \left(\kappa M \pi_{\rm rel}\right)^{1/2},
\qquad
\pi_{\rm rel} = {\rm au}\left(\frac{1}{D_{\rm L}}-\frac{1}{D_{\rm S}}\right).
\label{eq2}
\end{equation}
\hskip-4pt
Here, $\kappa = 4G/(c^2{\rm au}) \simeq 8.144~{\rm mas}~M_\odot^{-1}$, $M$ is the 
total lens mass, and $\dl$ and $\ds$ denote the distances to the lens and source, 
respectively.

\begin{figure}[t]
\includegraphics[width=\columnwidth]{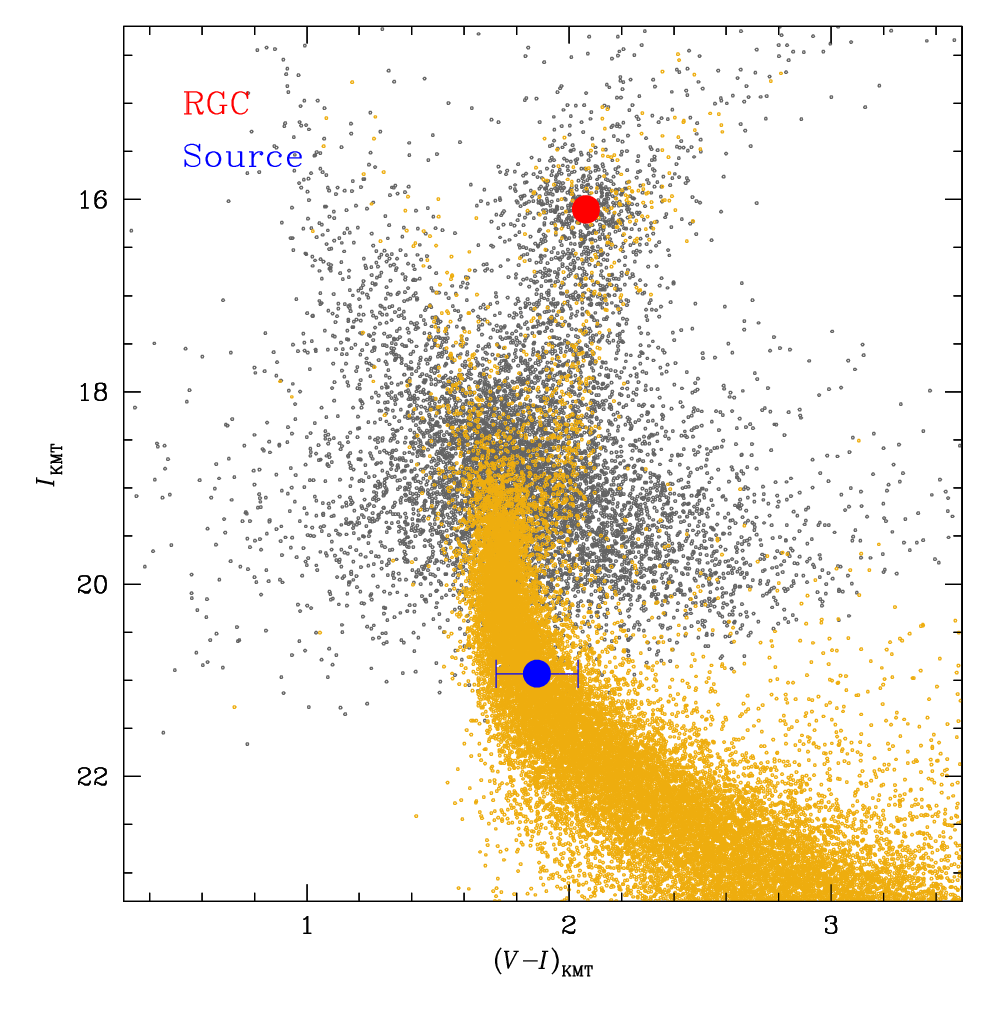}
\caption{
Source position in the instrumental color--magnitude diagram. Also marked is
the centroid of the red giant clump (RGC), which is used to calibrate the
source color and magnitude. The gray and brown points represent the CMDs
constructed from the KMTC and HST observations, respectively.
}
\label{fig:six}
\end{figure}

The source properties were determined from its position in the instrumental 
color--magnitude diagram (CMD) constructed using stars in the vicinity of the 
event.  We determined the instrumental color and magnitude of the source by 
regressing the $V$- and $I$-band photometry against the best-fit light-curve 
model.  For OGLE-2018-BLG-0829, however, it was difficult to measure the source 
color using the survey data because the $V$-band observations were both sparse 
and of low signal-to-noise ratio.  To overcome this limitation, we used the CMD 
constructed from \emph{Hubble Space Telescope}.  We first aligned the instrumental 
CMD obtained from the KMTC observations with the HST CMD using red giant clump 
(RGC) centroids in the two data sets. The source color was then estimated from 
the mean color of HST stars lying within the range of the measured source $I$-band 
magnitude.  Figure~\ref{fig:six} shows the source position in the combined HST 
and KMTC CMD.

\begin{deluxetable}{lllllll}
\tablewidth{0pt}
\tablecaption{Source parameters. \label{table:two}}
\tablehead{
\multicolumn{1}{c}{Parameter}         &
\multicolumn{1}{c}{Value    }   
}
\startdata
 $(V-I, I)_S            $   &  $(1.878 \pm 0.157, 20.930 \pm 0.069)$   \\
 $(V-I, I)_{\rm RGC}    $   &  $(2.065, 16.103)                    $   \\
 $(V-I, I)_{{\rm RGC},0}$   &  $(1.060, 14.426)                    $   \\
 $(V-I, I)_{S,0}        $   &  $(0.873 \pm 0.162, 19.253 \pm 0.072)$   \\
\enddata
\end{deluxetable}

The source color and magnitude were then calibrated relative to the RGC 
centroid using the method of \citet{Yoo2004}. Because the source and the 
RGC lie along nearly the same line of sight, they are expected to experience 
approximately the same extinction and reddening. The de-reddened source color 
and magnitude were therefore obtained from
\begin{equation}
(V-I, I)_{S,0} = (V-I, I)_{{\rm RGC},0} + \Delta(V-I, I),
\label{eq3}
\end{equation}
\hskip-4pt
where $\Delta(V-I, I) = (V-I, I)_S - (V-I, I)_{\rm RGC}$, $(V-I, I)_S$ and 
$(V-I, I)_{\rm RGC}$ are the positions of the source and the RGC centroid 
in the instrumental CMD, respectively, and $(V-I, I)_{{\rm RGC},0}$ is the 
known de-reddened position of the RGC centroid \citep{Bensby2013,Nataf2013}.

The uncertainties in the de-reddened source color and magnitude were estimated by 
adding the source measurement uncertainties in quadrature with the adopted uncertainties 
of 0.04 mag in the intrinsic RGC color and 0.02 mag in the intrinsic RGC magnitude.  
The source color was estimated from the mean color of HST stars within the magnitude 
range corresponding to the source brightness, and its uncertainty was determined from 
the standard deviation of the colors of the selected stars.  In estimating the angular 
source radius, we additionally adopted a 7\% fractional uncertainty to account for 
residual systematic uncertainties associated with the color--surface-brightness 
relation and photometric transformations. This contribution was added in quadrature 
to the uncertainty estimated from the de-reddened source color.

The measured and calibrated color and magnitude of the source, together with 
those of the RGC centroid, are listed in Table~\ref{table:two}. The calibrated 
color and magnitude indicate that the source is an early K-type main-sequence 
star located in the Galactic bulge.

Using the estimated de-reddened source color and magnitude, we determined the angular 
source radius from the empirical color--surface-brightness relation of \citet{Adams2018}.  
Applying this relation yields 
\begin{equation}
\theta_* = 0.525 \pm 0.093\,\mu{\rm as}.
\label{eq4}
\end{equation}
\hskip-4pt
Combining the measured angular source radius with the normalized source radius 
using Equation~(\ref{eq1}), we obtained the angular Einstein radius,
\begin{equation}
\thetae = 0.325 \pm 0.059\,{\rm mas}.
\label{eq5}
\end{equation}
\hskip-4pt
Finally, the relative lens--source proper motion was estimated from the angular 
Einstein radius and the event timescale,
\begin{equation}
\mu = \frac{\thetae}{\te} = 16.5 \pm 3.1\,{\rm mas}\,{\rm yr}^{-1}.
\label{eq6}
\end{equation}

\section{Physical lens parameters \label{sec:seven}}

The physical parameters of the lens cannot be uniquely determined from the light 
curve alone because the microlens-parallax parameter, $\pie$, was not measured. 
We therefore estimated the properties of the lens system by conducting a Bayesian 
analysis using the constraints provided by the measured event timescale, $\te$, 
and angular Einstein radius, $\thetae$.

For the Bayesian analysis, we generated a large ensemble of artificial microlensing 
events based on a Galactic model incorporating the spatial distributions, kinematics, 
and mass functions of disk and bulge lens populations.  We adopted the Galactic 
model of \citet{Jung2021} together with the lens mass function of \citet{Jung2022}. 
For each simulated event, we computed the lensing observables, $t_{{\rm E},i}$ and 
$\theta_{{\rm E},i}$, from the simulated lens mass, distance, and transverse velocity, 
and assigned a likelihood weight
\begin{equation}
w_i = \exp\left\{
-\frac{1}{2}
\left[
\left( \frac{t_{{\rm E},i}-t_{\rm E}}{\sigma(t_{\rm E})} \right)^2
+
\left( \frac{\theta_{{\rm E},i}-\theta_{\rm E}}
{\sigma(\theta_{\rm E})} \right)^2
\right]
\right\}.
\end{equation}
\hskip-4pt
The posterior distributions of the physical lens parameters were then
constructed from the weighted ensemble of simulated events.

From these posterior distributions, we estimated the primary-lens mass, $M_1$, 
and lens distance, $\dl$. For each parameter, we adopted the median of its 
posterior distribution as the representative value and the 16th and 84th 
percentiles as the lower and upper bounds of the uncertainty range, respectively. 
The masses of the two companions were subsequently derived using the measured 
mass ratios,
\begin{equation}
M_2 = q_2 M_1, \qquad M_3 = q_3 M_1.
\label{eq8}
\end{equation}
\hskip-4pt
The projected separations of the companions from the primary were calculated as
\begin{equation}
a_{\perp,2}=s_2 \dl\thetae, \qquad
a_{\perp,3}=s_3 \dl\thetae.
\label{eq9}
\end{equation}

\begin{figure}[t]
\includegraphics[width=\columnwidth]{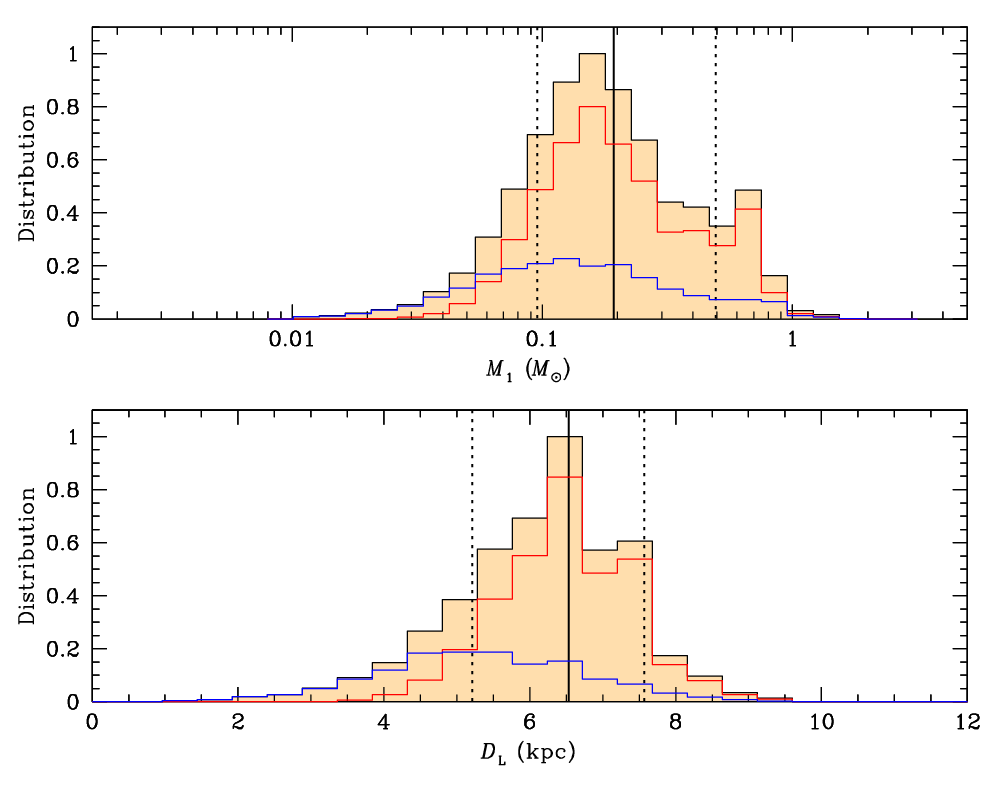}
\caption{
Bayesian posterior distributions of the host-star mass, $M_1$ (upper panel),
and the lens distance, $D_{\rm L}$ (lower panel). The black histograms show
the total posterior distributions, while the red and blue histograms represent
the contributions from Galactic bulge and disk lenses, respectively. 
For each distribution, the solid vertical line indicates the median (50th 
percentile), while the two dotted vertical lines mark the 16th and 84th 
percentiles of the total posterior distribution.
}
\label{fig:seven}
\end{figure}

Figure~\ref{fig:seven} presents the Bayesian posterior distributions of the 
host-star mass, $M_1$ (upper panel), and the lens distance, $D_{\rm L}$ (lower 
panel).  The black histograms represent the total posterior distributions, 
while the red and blue histograms show the contributions from Galactic bulge 
and disk lenses, respectively.  In each panel, the three vertical lines mark 
the 16th, 50th, and 84th percentiles of the total posterior distribution, with 
the solid line indicating the median.

The posterior mass distributions of the secondary and tertiary lens components 
are not shown separately because their masses are related to the primary mass 
by $M_2=q_2M_1$ and $M_3=q_3M_1$, respectively. Given the relatively small 
uncertainties in the mass ratios, their posterior distributions in logarithmic 
mass are essentially shifted versions of the primary mass distribution, with 
offsets of $\log q_2$ and $\log q_3$, respectively.

The estimated masses of the individual lens components, the lens distance, and 
the projected separations are summarized in Table~\ref{table:three}. The quoted 
values correspond to the medians of the posterior distributions, while the 
uncertainties represent the 16th and 84th percentiles. The table also lists the 
relative probabilities that the lens lies in the Galactic disk ($p_{\rm disk}$) 
or bulge ($p_{\rm bulge}$).

Bases on the median values, the Bayesian estimates indicate that the primary lens, 
$M_1$, is a low-mass star, while its binary companion, $M_2$, lies in the brown-dwarf 
mass regime.  Because of the large range of uncertainties, $M_1$ has a probability of 
brown dwarf of $\sim 14\%$ with a joint probability that both $M_1$ and $M_2$ are 
brown dwarfs is $\sim 11\%$.  The third component, $M_3$, has a mass in the planetary 
regime, approximately comparable to that of Saturn.  The brown-dwarf companion is 
located at a projected separation of $a_{\perp,2}= 3.58^{+0.57}_{-0.73}$\,au from 
the primary, whereas the planet lies at a projected separation of $a_{\perp,3}=
1.56^{+0.25}_{-0.32}$\,au.

\begin{deluxetable}{lllllll}
\tablewidth{0pt}
\tablecaption{Physical lens parameters estimated through Bayesian analysis. \label{table:three}}
\tablehead{
\multicolumn{1}{c}{Parameter}         &
\multicolumn{1}{c}{Value    }   
}
\startdata
  $M_1$ ($M_\odot$)    &  \hspace{2cm} $0.193^{+0.300}_{-0.098}$  \hspace{2cm} \\    [0.6ex]
  $M_2$ ($M_\odot$)    &  \hspace{2cm} $0.068^{+0.105}_{-0.034}$  \hspace{2cm} \\    [0.6ex]
  $M_3$ ($M_{\rm J}$)  &  \hspace{2cm} $0.337^{+0.522}_{-0.171}$  \hspace{2cm} \\    [0.6ex]
  $\dl$ (kpc)          &  \hspace{2cm} $6.54^{+1.03}_{-1.33}   $  \hspace{2cm} \\    [0.6ex]
  $a_{\perp,2}$ (au)   &  \hspace{2cm} $3.58^{+0.57}_{-0.73}   $  \hspace{2cm} \\    [0.6ex]
  $a_{\perp,3}$ (au)   &  \hspace{2cm} $1.56^{+0.25}_{-0.32}   $  \hspace{2cm} \\    [0.6ex]
  $p_{\rm disk}$       &  \hspace{2cm}  29\%                      \hspace{2cm} \\    [0.6ex]
  $p_{\rm bulge}$      &  \hspace{2cm}  71\%                      \hspace{2cm} \\    [0.6ex]
\enddata
\end{deluxetable}

The relatively high lens--source proper motion, $\mu = 16.5 \pm 3.1\,{\rm mas}\,{\rm yr}^{-1}$, 
favors a nearby or kinematically fast lens. Nevertheless, because neither the 
microlens parallax nor the lens flux was measured, the individual masses and 
distance remain subject to substantial uncertainties. Future high-angular-resolution 
imaging may resolve the lens from the source and thereby measure the lens flux and 
relative proper motion independently, providing a direct test of the inferred 
star--brown dwarf--planet interpretation.

\section{Discussion \label{sec:eight}}

We analyzed the microlensing event OGLE-2018-BLG-0829, whose light curve exhibits a 
weak anomaly in addition to prominent caustic-crossing features characteristic of a 
binary lens. Although the overall light-curve morphology is well reproduced by a 2L1S 
model, the weak anomaly preceding the first caustic crossing cannot be explained within 
this framework. We therefore explored both 2L2S and 3L1S interpretations and found 
that both models can reproduce the additional anomaly. 
However, the 2L2S model implies an unusual combination of source properties 
and is statistically disfavored relative to the 3L1S model by $\Delta\chi^2=15.1$. 
We therefore regard the 3L1S interpretation as the more plausible explanation among 
the models considered.

Under the favored 3L1S interpretation, the lens may be a hierarchical system 
consisting of a low-mass M dwarf, a brown-dwarf candidate, and a Saturn-mass 
planet. Bayesian analysis suggests that the two companions lie at projected 
separations of approximately 3.6 au and 1.6 au from the primary, respectively. 
However, the inferred component masses and orbital configuration remain uncertain, 
as discussed below. If this interpretation is confirmed, OGLE-2018-BLG-0829L 
would represent the third star--brown dwarf--planet system identified through 
microlensing.

The potential significance of this system lies in the insight it may provide into 
planet formation and survival in star--brown dwarf environments. Together with 
KMT-2020-BLG-0414Lb \citep{Zang2021} and KMT-2024-BLG-0404L \citep{Han2025}, 
OGLE-2018-BLG-0829L would contribute to a small but diverse sample of planetary 
systems containing brown-dwarf companions. These three systems would span nearly 
three orders of magnitude in planetary mass, ranging from an Earth-mass planet to 
a Saturn-mass planet. Although the sample remains far too small to support statistical 
conclusions, the diversity of these systems suggests that planet formation may occur 
under a range of conditions in systems containing brown-dwarf companions, motivating 
further observational and theoretical investigations.

Beyond the possible identification of a planet in a binary system, this event is 
also notable from a methodological perspective. The planetary signal manifests as 
a subtle deviation in the light curve superposed on the prominent caustic-crossing 
features produced by the underlying binary lens. Although the planet substantially 
modifies the caustic structure, its photometric signature is relatively weak and 
would likely have remained undetected without the high-precision re-reduction of 
the KMTNet data. This demonstrates that improved photometric precision can reveal 
planetary signals hidden in apparently well-understood binary-lens events. As 
photometric reduction techniques continue to improve and increasingly large 
microlensing data sets become available, similar weak perturbations may reveal 
additional planets in complex multiple-lens systems.

The relatively high lens--source relative proper motion makes this event 
a promising target for high-angular-resolution follow-up observations. With a 
measured relative proper motion of $\mu \sim 17$ mas yr$^{-1}$, the lens and 
source are expected to be separated by approximately 130 mas eight years after 
the event, potentially allowing them to be resolved with current adaptive-optics 
facilities such as Keck. A direct measurement of the lens flux would provide 
additional constraints on the mass and distance of the luminous lens component, 
thereby offering an independent test of the physical parameters inferred under 
the favored 3L1S interpretation.

\section{Limitations of the Current Analysis \label{sec:nine}}

Although the 3L1S model provides the best fit among the models considered, the 
present analysis should be regarded as a preliminary exploration of plausible 
interpretations rather than a definitive characterization of the lens system. 
Several effects not included in the modeling could influence the inferred lens 
parameters and potentially alter the preferred interpretation.

First, the modeling did not include microlens parallax \citep{Gould1992a,An2002} 
or orbital motion of the lens components \citep{Dominik1998,Albrow2000}. Although 
the relatively short event timescale may limit the sensitivity to parallax, the 
presence of multiple sharp light-curve features could provide additional constraints 
on this effect. Similarly, orbital motion could modify the caustic geometry during 
the event and affect the relative goodness of fit of the competing solutions. 
Including these effects could therefore change the best-fit parameters and increase 
their uncertainties.

Second, the substantial blending may introduce a degeneracy between the blending 
flux and the Einstein timescale, potentially increasing the uncertainties in the 
inferred lens parameters \citep{Han1997,Han2000}. If a significant fraction of 
the blended light originates from the lens system, its brightness could provide an 
additional constraint on the lens mass and distance, thereby affecting the Bayesian 
interpretation. An astrometric analysis of the blended light, together with future 
high-angular-resolution observations, could help determine its origin and refine 
the inferred physical properties of the lens system.

Another important consideration is the dynamical stability of the proposed system. 
The inferred projected separations of the brown-dwarf candidate and the planet are 
approximately 3.6 au and 1.6 au, respectively, corresponding to a separation ratio 
of only $\sim 2.3$. For a planet orbiting one component of a binary, long-term 
stability generally requires the planetary orbit to be substantially smaller than 
the binary orbit \citep{Holman1999}. The relatively small projected separation 
ratio therefore raises questions about the dynamical stability of the proposed 
configuration. However, the projected separations do not directly determine the 
orbital semi-major axes, which depend on the unknown orbital orientations, phases, 
and eccentricities. A detailed dynamical analysis would therefore be required to 
assess the long-term stability of the proposed system.

In addition, although the favored 3L1S model reproduces the principal features of 
the observed light curve, small residuals remain around the planetary anomaly. These 
residuals may indicate limitations of the static-lens approximation or the presence 
of alternative solutions that were not identified in the present search. More 
extensive modeling, incorporating additional physical effects and a broader 
exploration of parameter space, would be needed to establish the uniqueness and 
physical plausibility of the favored interpretation.

Taken together, these considerations indicate that the proposed star--brown dwarf--planet 
configuration remains tentative. Nevertheless, the present analysis identifies a 
plausible interpretation of the observed anomaly and provides a basis for further 
investigation.

\begin{acknowledgements}
C.H. was supported by the National Research Foundation grant (RS-2025-21073000) 
funded by the Korea government.
This research has made use of the KMTNet system operated by the Korea Astronomy and Space 
Science Institute (KASI) at three host sites of CTIO in Chile, SAAO in South Africa, and 
SSO in Australia. Data transfer from the host site to KASI was supported by the Korea 
Research Environment Open NETwork (KREONET). This research was supported by KASI under 
the R\&D program (project No. 2026-1-904-01) supervised by the Ministry of Science and ICT.
Q.Q., W.Z., H.Y., H.L., J.Z. and S.M. acknowledge support by the National Natural Science
Foundation of China (Grant No. 12133005).
The OGLE project has received funding from the Polish National Science
Centre grant OPUS-28 2024/55/B/ST9/00447 awarded to AU.
\end{acknowledgements}


\bibliography{references}

@article{Adams2018,
  author  = {Adams, A. D. and Boyajian, T. S. and von Braun, K.},
  year    = {2018},
  title   = {Predicting Stellar Angular Sizes},
  journal = {\mnras},
  volume  = {473},
  pages   = {3608--3614},
  doi     = {10.1093/mnras/stx2367}
}

@article{Albrow2000,
  author  = {Albrow, M. D. and Beaulieu, J.-P. and
             Caldwell, J. A. R. and Dominik, M. and
             Gaudi, B. S. and Gould, A. and others},
  year    = {2000},
  title   = {Detection of Rotation in a Binary Microlens:
             PLANET Photometry of MACHO 97-BLG-41},
  journal = {\apj},
  volume  = {534},
  pages   = {894--906},
  doi     = {10.1086/308798}
}

@article{An2002,
  author  = {An, J. H. and Albrow, M. D. and Beaulieu, J.-P.
             and Caldwell, J. A. R. and DePoy, D. L.
             and Dominik, M. and others},
  year    = {2002},
  title   = {First Microlens Mass Measurement:
             PLANET Photometry of EROS BLG-2000-5},
  journal = {\apj},
  volume  = {572},
  pages   = {521--539},
  doi     = {10.1086/340191}
}

@article{An2005,
  author = {An, J. H.},
  title = {Gravitational lens under perturbations: symmetry of perturbing potentials with invariant caustics}, 
  journal = {\mnras},
  volume = {356},
  pages = {1409--1416},
  year = {2005},
  doi = {10.1111/j.1365-2966.2004.08559.x}
}

@article{Artymowicz1994,
author = {Artymowicz, P. and Lubow, S.~H.},
title = {Dynamics of Binary-Disk Interaction. I. Resonances and Disk Gap Sizes},
journal = {\apj},
year = {1994},
volume = {421},
pages = {651--667},
doi = {10.1086/173679}
}

@article{Beaulieu2006,
author = {Beaulieu, J.-P. and Bennett, D. P. and Fouqu{\'e}, P. and
Williams, A. and Dominik, M. and J{\o}rgensen, U. G. and
Kubas, D. and Cassan, A. and Coutures, C. and Greenhill, J. and et al.},
title = {Discovery of a Cool Planet of 5.5 Earth Masses through Gravitational Microlensing},
journal = {Nature},
year = {2006},
volume = {439},
pages = {437--440},
doi = {10.1038/nature04441}
}

@article{Bennett2014,
author = {Bennett, David P. and Batista, Virginie and Bond, Ian A. and
Suzuki, Daisuke and Beaulieu, Jean-Philippe and Fukui, Akihiko and
Furusawa, Kunihiko and Hearnshaw, John B. and Itow, Yoshitaka and
Kilmartin, Peter M. and et al.},
title = {MOA-2010-BLG-073Lb: An M-Dwarf with a Substellar Companion at the
Planet/Brown-Dwarf Boundary},
journal = {\apj},
year = {2014},
volume = {785},
pages = {155},
doi = {10.1088/0004-637X/785/2/155}
}

@ARTICLE{Bensby2013,
author = {Bensby, T. and Yee, J.~C. and Feltzing, S. and Johnson, J.~A. and Gould, A. and
Ad{\'e}n, D. and Asplund, M. and Mel{\'e}ndez, J. and Ramirez, I. and others},
title = {Chemical evolution of the Galactic bulge as traced by microlensed dwarf and
subgiant stars. IV. Two bulge populations},
journal = {\aap},
year = {2013},
volume = {549},
pages = {A147},
doi = {10.1051/0004-6361/201220678}
}

@article{Bond2004,
author = {Bond, Ian A. and Udalski, Andrzej and Jaroszy{\'n}ski, Micha{\l} and
Rattenbury, Nicholas J. and Paczy{\'n}ski, Bohdan and
Soszy{\'n}ski, Igor and Wyrzykowski, {\L}ukasz and
Szyma{\'n}ski, Micha{\l} K. and Kubiak, Marek and
Szewczyk, Olga and et al.},
title = {OGLE 2003-BLG-235/MOA 2003-BLG-53: A Planetary Microlensing Event},
journal = {\apj},
year = {2004},
volume = {606},
pages = {L155--L158},
doi = {10.1086/420928}
}

@article{Bozza2000,
author = {Bozza, Valerio},
title = {Caustics in Special Multiple Lenses},
journal = {\aap},
year = {2000},
volume = {355},
pages = {423—432},
doi = {10.48550/arXiv.astro-ph/9910535}
}

@article{Danek2015,
author = {Dan{\v e}k, Karel and Heyrovsk{\'y}, David},
title = {Critical Curves and Caustics of Triple-lens Models},
journal = {\apj},
year = {2015},
volume = {806},
pages = {99},
doi = {10.1088/0004-637X/806/1/99}
}

@article{Danek2019,
author = {Dan{\v e}k, Karel and Heyrovsk{\'y}, David},
title = {Triple-lens Gravitational Microlensing: Critical Curves for Arbitrary Spatial Configuration},
journal = {\apj},
year = {2019},
volume = {880},
pages = {72},
doi = {10.3847/1538-4357/ab2982}
}

@article{Dominik1998,
  author  = {Dominik, M.},
  year    = {1998},
  title   = {Galactic Microlensing with Rotating Binaries},
  journal = {\aap},
  volume  = {329},
  pages   = {361--374}
}

@article{Dominik1999,
  author = {Dominik, Martin},
  title = {The Binary Gravitational Lens and Its Extreme Cases},
  journal = {\aap},
  volume = {349},
  pages = {108--125},
  year = {1999}
}

@article{Duquennoy1991,
author = {Duquennoy, A. and Mayor, M.},
title = {Multiplicity among solar-type stars in the solar neighbourhood. II. Distribution of the
orbital elements in an unbiased sample},
journal = {\aap},
volume = {248},
pages = {485--524},
year = {1991}
}

@article{Dominik2004,
  author  = {Dominik, M.},
  year    = {2004},
  title   = {Theory and Practice of Microlensing Light Curves around Fold Singularities},
  journal = {\mnras},
  volume  = {353},
  pages   = {69},
  doi     = {10.1111/j.1365-2966.2004.08046.x}
}

@article{Gaudi2008,
author = {Gaudi, B. Scott and Bennett, David P. and Udalski, Andrzej and
Gould, Andrew and Christie, Grant W. and Maoz, Dan and
Dong, Subo and McCormick, John and Szyma{\'n}ski, Micha{\l} K. and
Tristram, Paul J. and et al.},
title = {Discovery of a Jupiter/Saturn Analog with Gravitational Microlensing},
journal = {Science},
year = {2008},
volume = {319},
pages = {927--930},
doi = {10.1126/science.1151947}
}

@article{Gaudi2012,
author = {Gaudi, B. Scott},
title = {Microlensing by Exoplanets},
journal = {\araa},
year = {2012},
volume = {50},
pages = {411--453},
doi = {10.1146/annurev-astro-081811-125518}
}

@article{Gould1992a,
  author  = {Gould, A.},
  year    = {1992},
  title   = {Extending the MACHO Search to ~10^6 Solar Masses},
  journal = {\apj},
  volume  = {392},
  pages   = {442--451},
  doi     = {10.1086/171443}
}

@article{Gould1992b,
author = {Gould, Andrew and Loeb, Abraham},
title = {Discovering Planetary Systems through Gravitational Microlenses},
journal = {\apj},
year = {1992},
volume = {396},
pages = {104--114},
doi = {10.1086/171700}
}

@article{Gould1994,
  author  = {Gould, A.},
  year    = {1994},
  title   = {Proper Motions of MACHOs},
  journal = {\apjl},
  volume  = {421},
  pages   = {L71},
  doi     = {10.1086/187190}
}

@article{Han1997,
  author  = {Han, C.},
  year    = {1997},
  title   = {Correction for Blending Problem in Gravitational
             Microlensing Events by Using Hubble Space Telescope},
  journal = {\apj},
  volume  = {490},
  pages   = {51},
  doi     = {10.1086/304877}
}

@article{Han2000,
  author  = {Han, C.},
  year    = {2000},
  title   = {On the Baseline Flux Determination of Microlensing
             Events Detectable with the Difference Image
             Analysis Method},
  journal = {\mnras},
  volume  = {312},
  pages   = {807--812},
  doi     = {10.1046/j.1365-8711.2000.03171.x}
}

@article{Han2001,
author = {Han, C. and Chang, H.-Y. and An, J.-H. and Chang, K.},
title = {Properties of Microlensing Light Curves Induced by Multiple-Planet Systems},
journal = {\mnras},
year = {2001},
volume = {328},
pages = {986--994},
doi = {10.1046/j.1365-8711.2001.04973.x},
}

@article{Han2017,
author = {Han, Cheongho and Udalski, Andrzej and Gould, Andrew and
Lee, Chung-Uk and Shvartzvald, Yossi and Zang, Weicheng and
Mao, Shude and Koz{\l}owski, Szymon and Albrow, Michael D. and
Chung, Sun-Ju and Hwang, Kyu-Ha and Jung, Youn Kil and
Kim, Dae-Il and Kim, Hyoun-Woo and Ryu, Yoon-Hyun and
Shin, In-Gu and Yee, Jennifer C. and Zhu, Wei and et al.},
title = {OGLE-2016-BLG-0613LABb: A Microlensing Planet in a Binary System},
journal = {\aj},
year = {2017},
volume = {154},
pages = {223},
doi = {10.3847/1538-3881/aa9179}
}

@article{Han2025,
  author       = {Han, Cheongho and Udalski, Andrzej and Lee, Chung-Uk and
                  Ryu, Yoon-Hyun and Albrow, Michael D. and Chung, Sun-Ju and
                  Gould, Andrew and Hwang, Kyu-Ha and Jung, Youn Kil and
                  Shin, In-Gu and Shvartzvald, Yossi and Yee, Jennifer C. and
                  Yang, Hongjing and Zang, Weicheng and Kim, Doeon and
                  Kim, Dong-Jin and Park, Byeong-Gon and Pogge, Richard W.
                  and others},
  title        = {KMT-2024-BLG-0404L: A Triple Microlensing System Consisting of a Star, a Brown Dwarf, and a Planet},
  journal      = {\aap},
  year         = {2025},
  volume       = {700},
  pages        = {A181},
  doi          = {10.1051/0004-6361/202452760}
}

@article{Han2026,
author = {Han, C. and Lee, C.-U. and Udalski, A. and Albrow, M. D. and
Chung, S.-J. and Gould, A. and Jung, Y. K. and Hwang, K.-H. and
Ryu, Y.-H. and Shvartzvald, Y. and Shin, I.-G. and Yee, J. C. and et al.},
title = {Four-body Gravitational Microlensing Events Involving Both a Binary Lens and a
Binary Source},
journal = {\apj},
year = {2026},
volume = {1004},
pages = {183},
doi = {10.3847/1538-4357/ae6ef2} 
}

@article{Holman1999,
  author  = {Holman, M. J. and Wiegert, P. A.},
  year    = {1999},
  title   = {Long-Term Stability of Planets in Binary Systems},
  journal = {\aj},
  volume  = {117},
  pages   = {621--628},
  doi     = {10.1086/300695}
}

@article{Jung2017,
author = {Jung, Y. K. and Udalski, A. and Bond, I. A. and Yee, J. C. and Gould, A. and
Albrow, M. D. and Chung, S.-J. and Han, C. and Hwang, K.-H. and
Kim, D. and Kim, D.-J. and Kim, S.-L. and Lee, C.-U. and
Park, B.-G. and Pogge, R. W. and Ryu, Y.-H. and Shvartzvald, Y. and
Shin, I.-G. and Szyma{\'n}ski, M. K. and Udalski, A. and
others},
title = {OGLE-2016-BLG-1003: First Resolved Caustic-crossing Binary-source Event Discovered
by Second-generation Microlensing Surveys},
journal = {\apj},
year = {2017},
volume = {841},
pages = {75},
doi = {10.3847/1538-4357/aa7057}
}

@article{Jung2021,
  author  = {Jung, Y. K. and Han, C. and Udalski, A. and Gould, A. and Yee, J. C. and Albrow, M. D. and Chung, S.-J. and Hwang, K.-H.},
  year    = {2021},
  title   = {OGLE-2018-BLG-0567Lb and OGLE-2018-BLG-0962Lb: Two Microlensing Planets through the Planetary-caustic Channel},
  journal = {\aj},              
  volume  = {161},
  pages   = {293},
  doi     = {10.3847/1538-3881/abf8bd}
}

@article{Jung2022,
  author  = {Jung, Y. K. and Zang, W. and Han, C. and Gould, A. and Udalski, A. and Albrow, M. D. and Chung, S.-J. and Hwang, K.-H.},
  year    = {2022},
  title   = {Systematic KMTNet Planetary Anomaly Search. VI. Complete Sample of 2018 Sub-prime-field Planets},
  journal = {\aj},              
  volume  = {164},
  pages   = {262},
  doi     = {10.3847/1538-3881/ac9c5c}
}

@article{Kim2016,
  author  = {Kim, S.-L. and Lee, C.-U. and Park, B.-G. and
             Kim, D.-J. and Cha, S.-M. and Lee, Y. and
             Han, C. and Chun, M.-Y. and Yuk, I. and others},
  year    = {2016},
  title   = {{KMTNet}: A Network of 1.6 m Wide-Field Optical
             Telescopes Installed at Three Southern Observatories},
  journal = {JKAS},
  volume  = {49},
  pages   = {37--44},
  doi     = {10.5303/JKAS.2016.49.1.37}
}

@article{Kraus2012,
author = {Kraus, A. L. and Ireland, M. J. and Hillenbrand, L. A. and Martinache, F.},
title = {The Role of Multiplicity in Disk Evolution and Planet Formation},
journal = {\apj},
year = {2012},
volume = {745},
pages = {19},
doi = {10.1088/0004-637X/745/1/19}
}

@article{Lee2008,
author = {Lee, Dong-Wook and Lee, Chung-Uk and Park, Byeong-Gon and
Chung, Sun-Ju and Kim, Young-Soo and Kim, Ho-Il and Han, Cheongho},
title = {Microlensing Detections of Planets in Binary Stellar Systems},
journal = {\apj},
year = {2008},
volume = {672},
pages = {623--629},
doi = {10.1086/523662}
}

@article{Luhman2012,
author = {Luhman, Kevin L.},
title = {The Formation and Early Evolution of Low-Mass Stars and Brown Dwarfs},
journal = {\araa},
year = {2012},
volume = {50},
pages = {65--106},
doi = {10.1146/annurev-astro-081811-125528}
}

@article{Mao1991,
author = {Mao, Shude and Paczy{\'n}ski, Bohdan},
title = {Gravitational Microlensing by Double Stars and Planetary Systems},
journal = {\apjl},
year = {1991},
volume = {374},
pages = {L37--L40},
doi = {10.1086/186066}
}

@article{Mroz2017,
author = {Mr{\'o}z, Przemek and Udalski, Andrzej and Skowron, Jan and
Poleski, Rados{\l}aw and Koz{\l}owski, Szymon and
Szyma{\'n}ski, Micha{\l} K. and Soszy{\'n}ski, Igor and
Wyrzykowski, {\L}ukasz and Pietrukowicz, Pawe{\l} and
Ulaczyk, Krzysztof and Skowron, Dorota and Pawlak, Micha{\l}},
title = {No Large Population of Unbound or Wide-orbit Jupiter-mass Planets},
journal = {Nature},
year = {2017},
volume = {548},
pages = {183--186},
doi = {10.1038/nature23276}
}

@ARTICLE{Nataf2013,
author = {Nataf, D. M. and Gould, A. and Fouqu{\'e}, P. and Gonzalez, O. A. and
Johnson, J. A. and Skowron, J. and Udalski, A. and Beaulieu, J.-P. and others},
title = {Reddening and extinction toward the Galactic bulge from OGLE-III: The
inner Milky Way's RV ≈ 2.5 extinction curve},
journal = {\apj},
year = {2013},
volume = {769},
pages = {88},
doi = {10.1088/0004-637X/769/2/88}
}

@article{Raghavan2010,
author = {Raghavan, D. and McAlister, H.~A. and Henry, T.~J. and
Latham, D.~W. and Marcy, G.~W. and Mason, B.~D. and
Gies, D.~R. and White, R.~J. and ten Brummelaar, T.~A.},
title = {A Survey of Stellar Families: Multiplicity of Solar-type Stars},
journal = {\apjs},
year = {2010},
volume = {190},
pages = {1--42},
doi = {10.1088/0067-0049/190/1/1}
}

@article{Schneider1987,
  author  = {Schneider, P. and Weiss, A.},
  year    = {1987},
  title   = {A Gravitational Lens Origin for AGN-Variability?},
  journal = {\aap},
  volume  = {171},
  pages   = {49}
}

@article{Udalski2015,
  author  = {Udalski, A. and Szyma{\'n}ski, M. K. and Szyma{\'n}ski, G.},
  year    = {2015},
  title   = {OGLE-IV: Fourth Phase of the Optical Gravitational Lensing Experiment},
  journal = {Acta Astron.},
  volume  = {65},
  pages   = {1},
  doi = {10.48550/arXiv.1504.05966}
}

@inproceedings{Whitworth2007,
author = {Whitworth, A. and Bate, M. R. and Nordlund, {\AA}ke and Reipurth, B.
and Zinnecker, H.},
title = {The Formation of Brown Dwarfs},
booktitle = {Protostars and Planets V},
editor = {Reipurth, B. and Jewitt, D. and Keil, K.},
pages = {459--476},
year = {2007},
publisher = {University of Arizona Press},
address = {Tucson}
}

@article{Yang2024,
  author       = {Yang, Hongjing and Yee, Jennifer C. and Hwang, Kyu-Ha and
                  Qian, Qiyue and Bond, Ian A. and Gould, Andrew and
                  Hu, Zhecheng and Zhang, Jiyuan and Mao, Shude and
                  Zhu, Wei and others},
  title        = {Systematic Reanalysis of KMTNet Microlensing Events, Paper I:
                  Updates of the Photometry Pipeline and a New Planet Candidate},
  journal      = {Monthly Notices of the Royal Astronomical Society},
  year         = {2024},
  volume       = {528},
  pages        = {11},
  doi          = {10.1093/mnras/stad3672}
}

@ARTICLE{Yoo2004,
author = {Yoo, J. and DePoy, D.~L. and Gal-Yam, A. and Gaudi, B.~S. and Han, C. and
Park, B.-G. and Pogge, R.~W. and Bennett, D.~P. and Udalski, A.},
title = {OGLE-2003-BLG-262: Finite-Source Effects from a Point-Mass Lens},
journal = {\apj},
year = {2004},
volume = {603},
pages = {139},
doi = {10.1086/381241}
}

@article{Zang2021,
  author       = {Zang, Weicheng and Han, Cheongho and Kondo, Iona and Yee, 
                  Jennifer C. and Lee, Chung-Uk and Gould, Andrew and Mao, Shude and
                  de Almeida, Leandro and Shvartzvald, Yossi and Zhang, Xiang-Yu
                  and others},
  title        = {An Earth-mass Planet in a Time of COVID-19: KMT-2020-BLG-0414Lb},
  journal      = {Research in Astronomy and Astrophysics},
  year         = {2021},
  volume       = {21},
  pages        = {239},
  doi          = {10.1088/1674-4527/21/9/239}
}
\bibliographystyle{aasjournalv7}

\end{document}